\documentclass{optica-article}

\journal{opticajournal} 

\articletype{Research Article}
\usepackage{amsmath}
\usepackage{subcaption}
\usepackage{graphicx}
\usepackage{lineno}

\usepackage{booktabs}
\usepackage{siunitx}

\usepackage{graphicx} 
\usepackage{xcolor}
\usepackage{color,soul}
\usepackage{float} 

\usepackage{subcaption}
\usepackage{upgreek}
\usepackage{svg}

\usepackage[most]{tcolorbox}
\usepackage{tikz}
\usepackage{adjustbox}

\begin{document}

\title{Bridging the Digital Divide: Small Language Models as a Pathway for Physics and Photonics Education in Underdeveloped Regions}

\author{Asghar Ghorbani,\authormark{1,}\authormark{*}}
\author{Hanieh Fattahi,\authormark{2,}\authormark{3,}\authormark{+}}
\authormark{1}LLM venture, Nuremberg, 90425, Germany

\authormark{2}Max Planck Institute for the Science of Light, Staudstrasse 2, Erlangen, 91058, Germany 

\authormark{3}Friedrich-Alexander-Universit{\"a}t Erlangen-N{\"u}rnberg, Staudstrasse 7, Erlangen, 91058, Germany

\email{\authormark{*}ghorbani59@gmail.com} 
\email{\authormark{+}hanieh.fattahi@mpl.mpg.de} 

\begin{abstract*} 
Limited infrastructure, scarce educational resources, and unreliable internet access often hinder physics and photonics education in underdeveloped regions. These barriers create deep inequities in Science, Technology, Engineering, and Mathematics (STEM) education. This article explores how Small Language Models (SLMs)—compact, AI-powered tools that can run offline on low-power devices, offering a scalable solution. By acting as virtual tutors, enabling native-language instruction, and supporting interactive learning, SLMs can help address the shortage of trained educators and laboratory access. By narrowing the digital divide through targeted investment in AI technologies, SLMs present a scalable and inclusive solution to advance STEM education and foster scientific empowerment in marginalized communities.
\end{abstract*}

\section{The Digital Divide in Education in underdeveloped regions}

\begin{figure}[th!]
    \centering
    \includegraphics[width=1\linewidth]{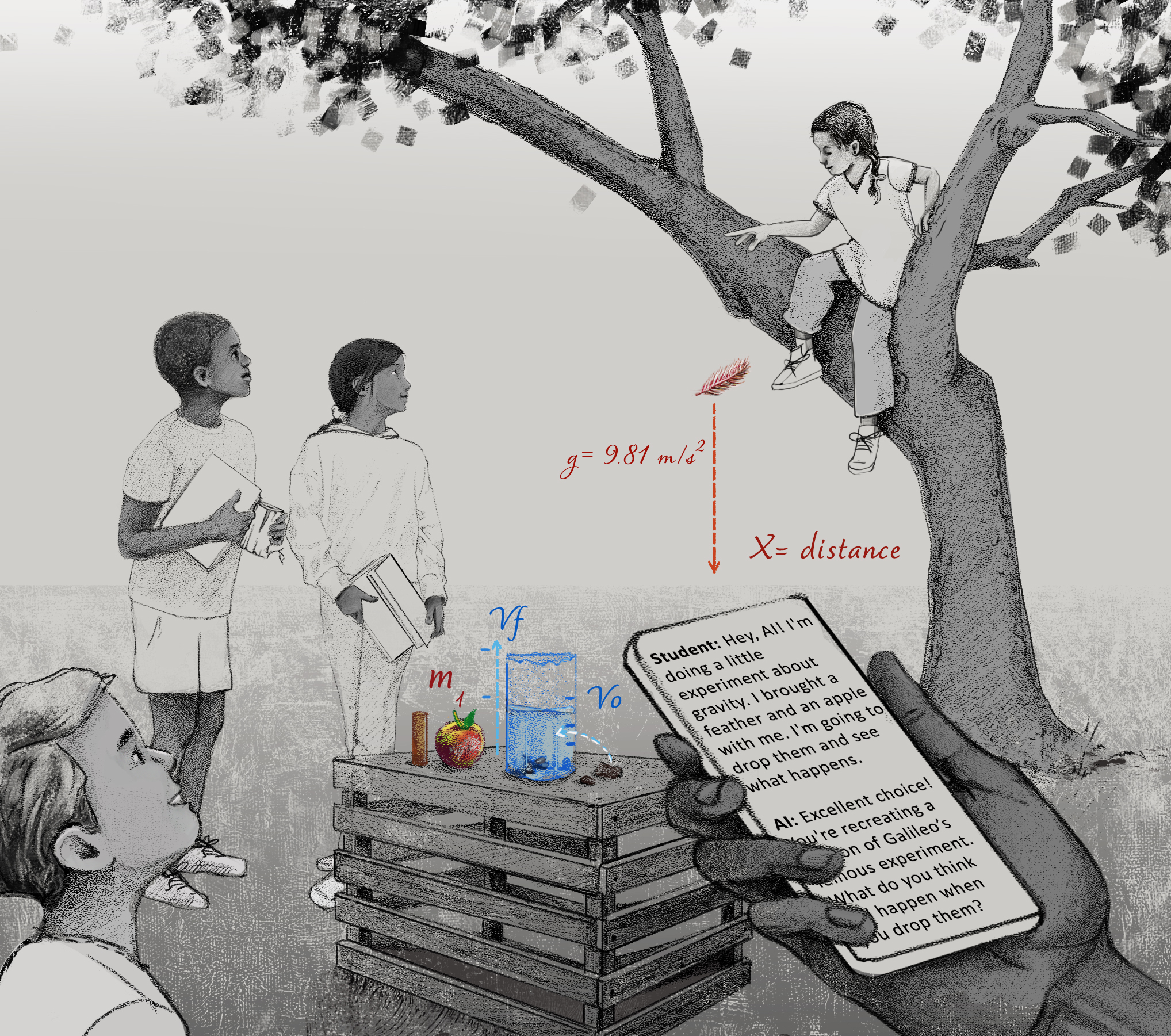}
    \caption{Small Language Models can help bridge the digital divide by enabling interactive on-site science education, bringing AI-powered learning to students without the need for internet access or advanced infrastructure. Illustration by S. Cook-Ordonez.}
    \label{fig:enter-label}
\end{figure}

Science, Technology, Engineering, and Mathematics (STEM) education is a driving force behind innovation, economic growth, and technological progress. It forms the backbone of advancements in critical fields such as medicine, engineering, and photonics, which are essential for sustainable development. Yet in many underdeveloped regions, access to quality STEM education remains deeply unequal, depriving students of the opportunity to contribute to and benefit from scientific and technological progress. The state of physics education in Africa exemplifies this disparity. According to a report by the African Development Bank \cite{AfDB2021}, fewer than 5\% of African students pursue tertiary education in STEM fields, with physics being one of the least chosen subjects.

Several factors contribute to the underrepresentation of students in STEM majors across underdeveloped countries. One of the most pressing issues is the lack of adequate educational infrastructure. Many schools in these regions lack the necessary resources to deliver effective STEM education. For instance, approximately 80\% of secondary schools in Africa do not have access to electricity, and over 90\% lack adequately equipped science laboratories. In countries such as Malawi and Tanzania, fewer than 15\% of schools have functional science laboratories, severely limiting hands-on learning opportunities \cite{WorldBank2020}. This shortfall not only inhibits students from conducting practical experiments, but also diminishes their interest and engagement in STEM subjects \cite{worldbank2024stem}.

Another major challenge is the shortage of qualified educators. Many teachers lack the specialized training necessary to teach STEM subjects effectively, leading to poor student performance and a lack of confidence in pursuing STEM careers. Furthermore, socioeconomic barriers significantly impact students' ability to enroll and persist in STEM programs. Poverty and economic hardship often force students to prioritize immediate income over education. Many families rely on their children to work instead of attending school, leading to lower enrollment and retention rates in STEM disciplines \cite{wikipediaEducationAfrica}.

Gender disparities further exacerbate the problem, as cultural norms and societal expectations in many underdeveloped countries discourage women from pursuing STEM careers. In Pakistan, for example, only 21\% of university students in engineering are women, and they make up merely 4.9\% of the engineering workforce \cite{wikipediaPakistaniWomenSTEM}. Overcoming these societal barriers requires targeted policies, mentorship programs, and initiatives that foster gender inclusivity in STEM education.

Despite increasing school enrollment, many students in developing countries fail to acquire fundamental literacy and numeracy skills, further limiting their ability to excel in STEM fields. In Kenya, Tanzania, and Uganda, 75\% of third-grade students cannot read an introductory sentence, highlighting a significant learning crisis that impedes their progression into more advanced STEM studies \cite{wikipediaLearningCrisis}. Without a strong foundational education, students struggle to grasp complex scientific and mathematical concepts, reducing their likelihood of pursuing careers in these critical fields. Addressing these challenges requires comprehensive educational reforms, investments in infrastructure, and targeted interventions to create equitable opportunities for students in underdeveloped regions. All these developments are crucial. However, they require significant investment in infrastructure and facilities, training programs for teachers, and cultural and economical advancement. 

Artificial intelligence (AI) has the potential to overcome these challenges with lower investment by enabling personalized on-site education. In recent years, AI and machine learning have transformed education across the world. Among these innovations, large language models (LLMs) such as OpenAI’s ChatGPT or DeepSeek-R1 have emerged as powerful tools for learning, assisting students and educators in gaining deeper insights and fostering creativity. However, their utility remains dependent mainly on stable Internet access, a luxury that is not available to many students and educators in underdeveloped countries. The digital divide and limited access to technology and the Internet in underdeveloped regions restrict students' exposure to digital tools and resources. Although mobile phone penetration has increased in recent years, reliable internet connectivity remains scarce in many rural and urban communities. According to UNESCO, approximately 89\% of students in sub-Saharan Africa lack access to household computers, and over 80\% lack access to the Internet \cite{united2020global}. This digital divide exacerbates existing educational inequities and leaves students without access to modern resources, such as AI-powered educational tools, that could otherwise transform learning outcomes in STEM fields like physics and photonics.
 
Addressing this disparity, Small Language Models (SLMs) offer a transformative solution for democratizing education in regions with limited Internet infrastructure. Unlike larger AI models that depend on cloud-based infrastructure and stable internet connectivity, SLMs are lightweight, domain-specific, and designed for local deployment. They can be optimized for STEM education (e.g., physics fundamentals) and run efficiently on low-end smartphones or computers, enabling interactive, personalized learning even in offline environments. This makes them uniquely suited to regions with unreliable internet access. This adaptability makes them well suited to tackling many real-world educational challenges, empowering learners in resource-constrained environments. In this article, we explore the state-of-the-art in SLMs and highlight their potential to advance scientific research and education in settings with limited connectivity.

\section{The Role and Limitations of Large Language Models (LLMs)}

LLMs power modern chatbots like OpenAI's ChatGPT, Anthropic's Claude, Google's Gemini, DeepSeek, and others, enabling them to process natural language inputs, known as prompts, and generate contextually relevant responses based on provided instructions. These models represent a paradigm shift in natural language processing, driven by \textit{transformer architectures} that address the critical limitations of earlier sequential models. A key concept in language modeling is the token, which refers to the fundamental units of text that the model processes. Tokens can be individual words, subword units, or even characters, depending on how the text is tokenized. For example, in a sentence like ``machine learning," the model might treat it as two separate tokens (``machine" and ``learning") or as a single unit, depending on the tokenization strategy \cite{huggingface_tokenizer_summary}. Traditional approaches, such as recurrent neural networks (RNNs) and convolutional neural networks (CNNs), processed tokens sequentially, which limited their ability to capture long-range dependencies. For instance, they frequently misinterpreted contextual relationships, such as associating pronouns with their correct antecedents or verbs with their subjects across extended text sequences~\cite{hochreiter97lstm, Lecun98cnn, vaswani2023attentionneed}.

A fundamental breakthrough in \textit{transformer architectures} is the \textit{self-attention} mechanism, which effectively addresses these limitations. Unlike previous models that processed tokens sequentially, \textit{self-attention} enables the model to compute relationships among all tokens simultaneously, regardless of their positions. This capability allows transformers to efficiently capture long-range dependencies, significantly enhancing their ability to understand complex linguistic patterns and improving performance across a wide range of language tasks \cite{bahdanau2016neuralmachinetranslationjointly, vaswani2023attentionneed}.

\begin{figure}[htbp]
    \centering
    \begin{subfigure}[b]{\textwidth}
        \centering
        \includegraphics[width=1.1\textwidth]{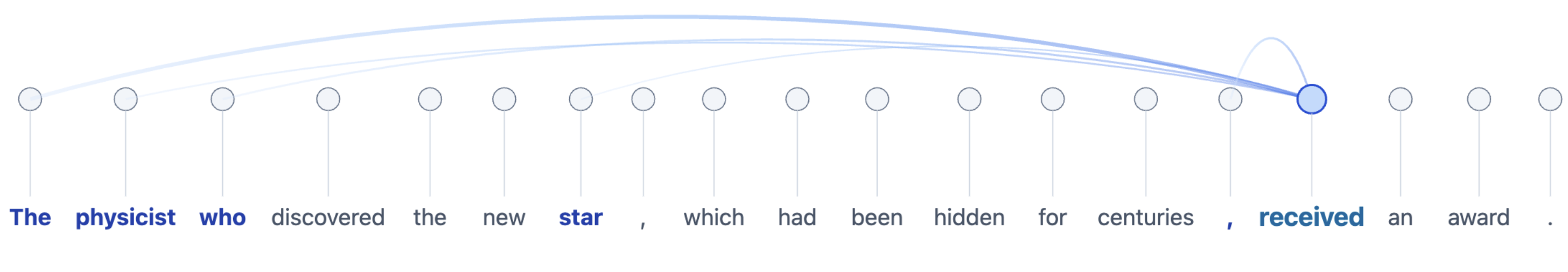}
        \caption{}
        \label{fig:attention-plot}
    \end{subfigure}
    
    \vspace{1cm}
    
    \begin{subfigure}[b]{\textwidth}
        \centering
        \includegraphics[width=1.1\textwidth]{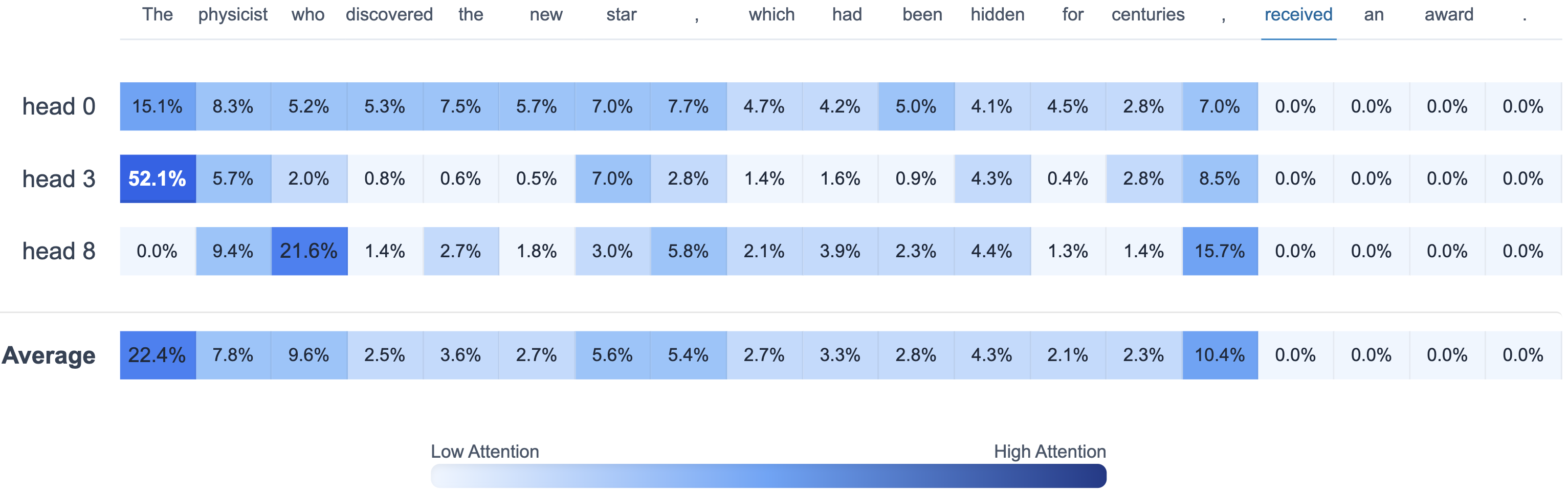}
        \caption{}
        \label{fig:attention-heatmap}
    \end{subfigure}
    \caption{Visualization of attention mechanisms in GPT-2 for the sentence: ``\textit{The physicist who discovered the new star, which had been hidden for centuries, received an award.}'' (a) Attention weights from the token ``received'' to preceding tokens in GPT-2. For clarity, only the top five highest-weighted connections are shown. (b) Attention heatmap displaying three selected heads (1, 3, and 8) from the final layer of GPT-2. Head 1 focuses on ``The physicist,'' Head 3 emphasizes ``The'' and punctuation, and Head 8 highlights ``who'' and punctuation. Values are normalized as percentages.}
    \label{fig:attn}
\end{figure}

Consider the following sentence: ``\textit{The physicist who discovered the new star, which had been hidden for centuries, received an award.}'' Here, \textit{self-attention} effectively would resolve dependencies across varying distances, including long-range associations that traditional models struggled with (Figure \ref{fig:attn}). For instance, the verb ``received'' (far from its subject) correctly links back to ``physicist'' despite intervening clauses (``who discovered... centuries'').

When generating ``received'', the self-attention model specifically focuses on ``physicist'' rather than distracting words like ``star'' or ``centuries''. Similarly, when resolving ``who'', attention directly links it to ``physicist''.

Self-attention is especially effective when processing long texts, such as answering questions based on an extensive research paper. In these scenarios, the model processes the entire text—both the questions and the document content simultaneously, allowing it to efficiently identify relevant information across the full context and establish meaningful connections between distant parts of the text.

Another crucial component of Transformers is Multi-Head Attention, which allows the model to analyze text from multiple perspectives simultaneously. Intuitively, each attention head specializes in different linguistic aspects---one may focus on syntax, another on semantic meaning, and another on contextual relationships. By integrating these diverse insights, the model develops a more comprehensive and nuanced understanding of the text. Figure \ref{fig:attn} illustrates how a self-attention model utilizes \textit{multiple attention heads}, each potentially capturing different aspects of a sentence's meaning. The attention values shown are normalized values derived from GPT-2. Such a multi-headed self-attention mechanism allows models to dynamically highlight multiple layers of context simultaneously, thereby enhancing their overall understanding capabilities. However, the standard Multi-Head Attention mechanism demands substantial memory storage. 

Although LLMs have transformed natural language processing, their real-world deployment faces significant challenges that limit accessibility and sustainability. The computational demands of transformer architectures require expensive GPU clusters for both training and inference, often compelling organizations to depend on cloud-based solutions. While this centralized approach offers convenience, it introduces several critical limitations. A key concern is the escalating energy consumption and infrastructure costs. For instance, pretraining even a moderate-sized model like the 70-billion-parameter Llama~2 emitted 291 tons of CO\textsubscript{2} equivalent, which is comparable to the annual emissions of 65 gasoline-powered cars \cite{touvron2023llama2openfoundation, epa_greenhouse_gas}. Scaling to larger architectures exacerbates this problem. For example, training the 405-billion-parameter Llama~3.1 required 30.84 million GPU hours, seventeen times higher than the energy expenditure of Llama~2 (see Table \ref{tab:llama2-emissions}) \cite{grattafiori2024llama3herdmodels}. Proprietary models such as GPT-4, which are estimated to contain over one trillion parameters, are expected to incur significantly higher environmental costs, although exact figures have not been publicly disclosed by their developers.

\begin{table}[h!]
    \centering
    \begin{tabular}{l S[table-format=7.0] S[table-format=3.0] S[table-format=6.2]}
    \toprule
    \textbf{Model Size} & \textbf{Time (GPU hours)} & \textbf{Power (W)} & \textbf{Carbon Emitted (tCO\textsubscript{2} eq)} \\
    \midrule
    7 B  & 184320  & 400 & 31.22  \\
    13B & 368640  & 400 & 62.44  \\
    34B & 1038336 & 350 & 153.90 \\
    70B & 1720320 & 400 & 291.42 \\
    \midrule
    \textbf{Total} & \textbf{3311616} &  & \textbf{539.00} \\
    \bottomrule
    \end{tabular}
    \caption{CO\textsubscript{2} emissions during pretraining for Llama 2 models. Time: total GPU time required to train each model. Power consumption: peak power capacity per GPU device
 \cite{touvron2023llama2openfoundation}.}
 \label{tab:llama2-emissions}
\end{table}

\begin{figure}
    \centering
    \includegraphics[width=0.9\linewidth]{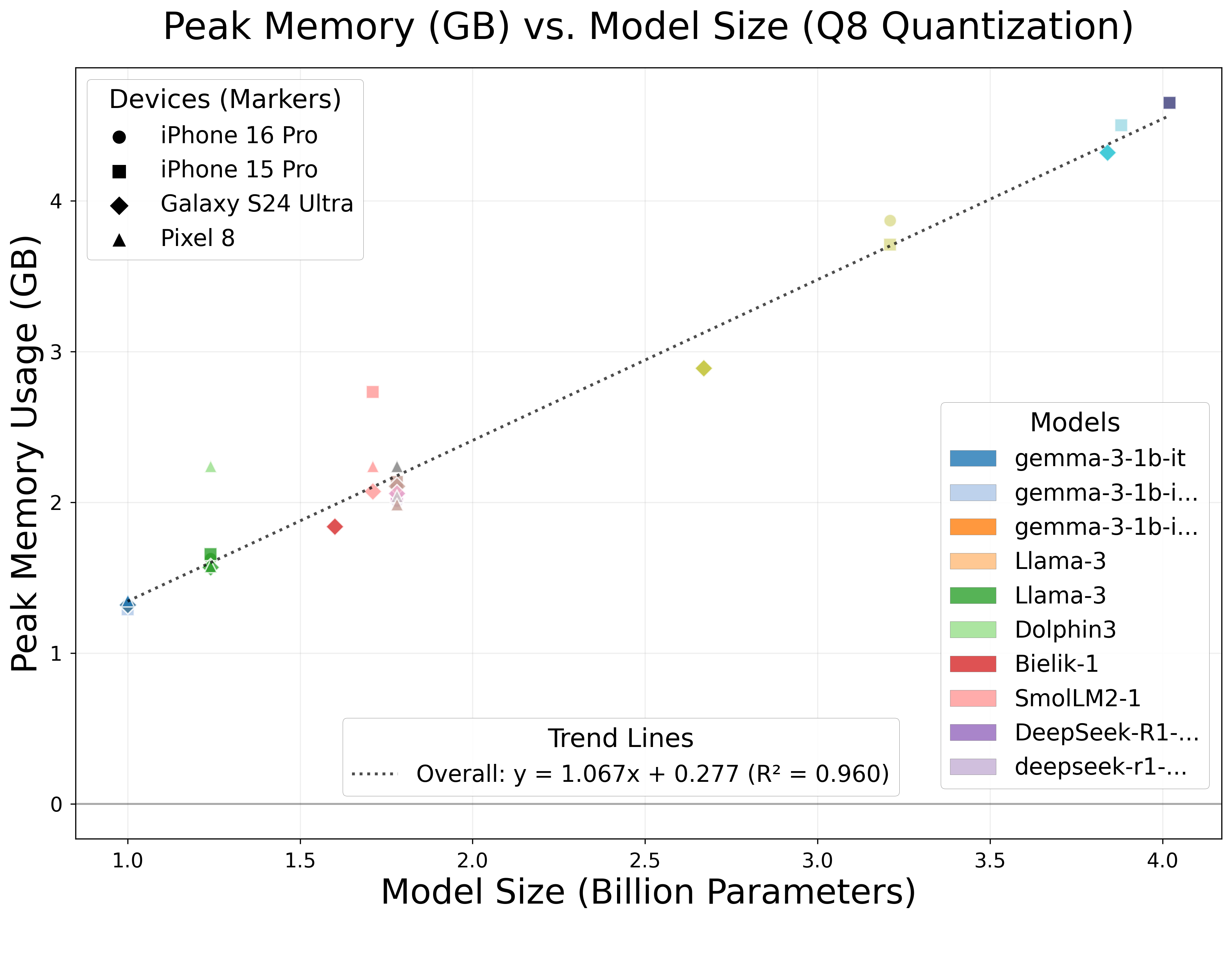}
    \caption{Peak memory consumption across devices for various model sizes under 8-bit quantization. Data sourced from the AI Phone Leaderboard benchmark \cite{huggingface-phone-leaderboard}.}
    \label{fig:memory-vs-modelsize}
\end{figure}

Training is only part of the story. From a deployment standpoint, a more pressing challenge lies in the resource demands of inference, particularly for real-time, on-device applications. Inference performance is primarily constrained by memory and compute availability. As illustrated in Figure~\ref{fig:memory-vs-modelsize}, peak memory usage during inference increases nearly linearly with model size under 8-bit quantization (Q8), where each parameter is represented using a single byte. Quantization helps lower both memory footprint and computational load by typically reducing parameter precision from 16 or 32 bits down to 8 bits. A more detailed discussion of this technique can be found in Section~\ref{sec:slm}. This near-linear scaling implies that a 70-billion-parameter model requires approximately 70\,GB of memory, which is far beyond the capacity of typical consumer devices. By contrast, mainstream smartphones generally offer only 2–6\,GB of memory, with high-end models rarely exceeding 10\,GB of available memory after accounting for system overhead. As a result, practical deployment is largely limited to models with roughly 4 billion parameters or fewer.

\begin{figure}
    \centering
    \includegraphics[width=0.9\linewidth]{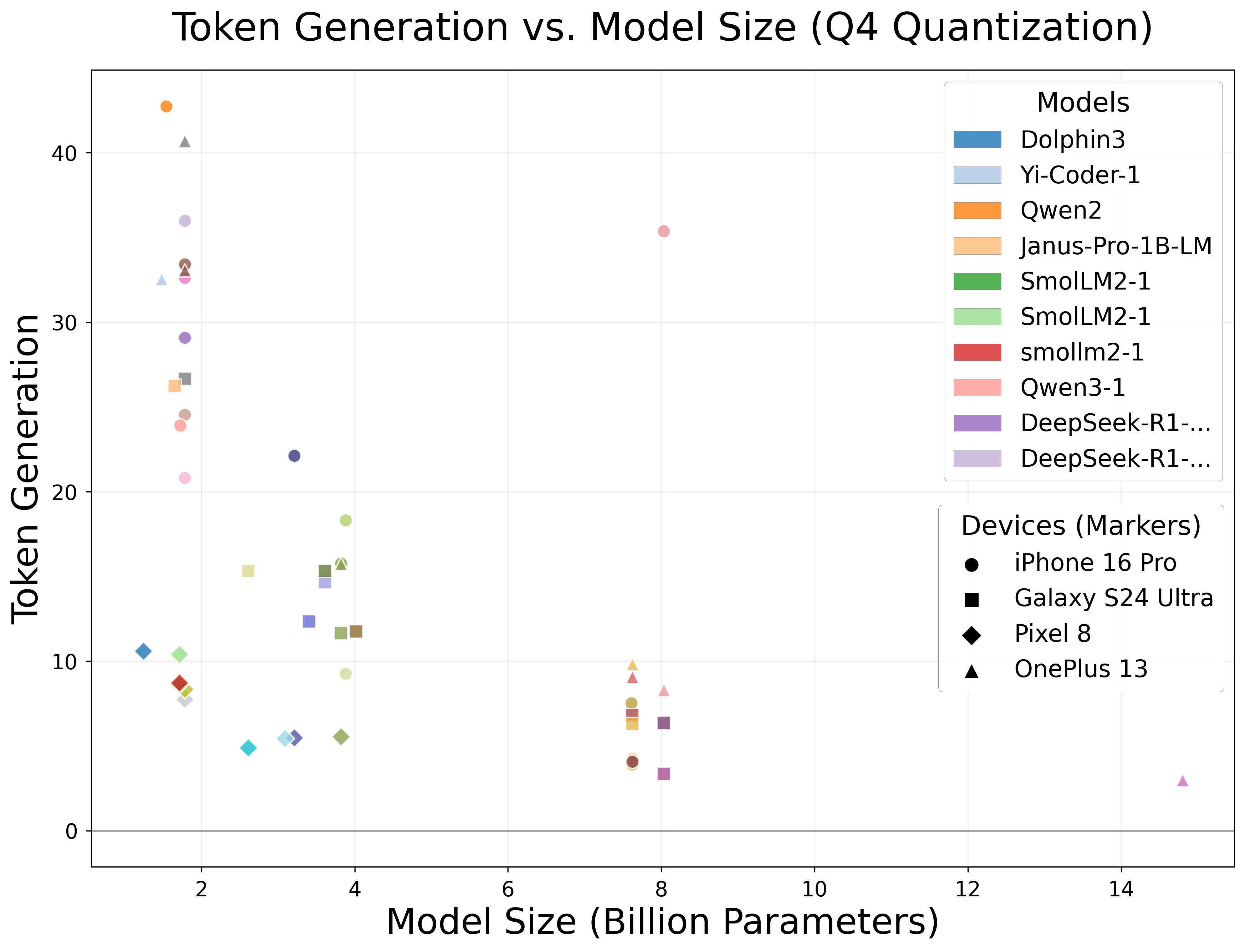}
    \caption{Token generation speed (tokens per second) across devices for various model sizes under 8-bit quantization. Data sourced from the AI Phone Leaderboard benchmark \cite{huggingface-phone-leaderboard}.}
    \label{fig:tokens-vs-modelsize}
\end{figure}

Additionally, inference latency becomes a critical bottleneck. As depicted in Figure~\ref{fig:tokens-vs-modelsize}, token generation speed (in tokens per second) declines sharply as model size increases, especially beyond the 4-billion-parameter threshold. This performance degradation hinders the responsiveness required for interactive applications, making large models unsuitable for real-time use on edge devices. Practically, a generation speed of around 10 tokens per second or higher is generally sufficient for many user-facing applications. However, when speeds drop below this threshold, the interaction becomes noticeably sluggish and may become impractical for real-time use. As illustrated in Figure~\ref{fig:tokens-vs-modelsize}, for models larger than 4 billion parameters, there are very few model-device combinations that maintain a generation speed above 10 tokens per second. These constraints highlight the need to optimize not just training efficiency, but also inference-time performance. They also underscore the growing interest in SLMs, which aim to strike a balance between capability and deployability on resource-constrained hardware.

Proprietary LLMs continue to push the boundaries of artificial intelligence, advancing toward artificial general intelligence and demonstrating exceptional performance in complex reasoning. However, their immense computational demands and dependence on cloud-based infrastructure limit their accessibility in many real-world applications. These models are particularly valuable in high-stakes research, advanced natural language processing, and enterprise AI solutions, where performance takes precedence over cost. Yet their prohibitive expenses and reliance on internet connectivity present significant barriers to widespread adoption, especially in settings that demand localized processing or offline functionality. To overcome these challenges, the AI community has increasingly shifted its focus toward SLMs, compact yet powerful architectures that strike a balance between performance, efficiency, and adaptability. Unlike their larger counterparts, SLMs emphasize accessibility and resilience, making them ideal for real-world applications where computational resources are constrained or connectivity is limited. As a result, SLMs represent a scalable, cost-effective alternative that supports inclusive and practical AI innovation across a wide range of environments.

\section{Small Language Models (SLMs): Efficiency, Adaptability, and Access}
\label{sec:slm}
SLMs have emerged as compelling alternatives to LLMs, designed to operate efficiently on resource-constrained devices while maintaining strong performance on domain-specific tasks \cite{allal2025smollm2smolgoesbig, grattafiori2024llama3herdmodels, liu2024mobilellmoptimizingsubbillionparameter, zhang2024tinyllamaopensourcesmalllanguage, qwen2025qwen25technicalreport, groeneveld2024olmoacceleratingsciencelanguage, pfeiffer2024h2odanube3technicalreport, microsoft2025phi4minitechnicalreportcompact, gemmateam2024gemmaopenmodelsbased}. While definitions of SLMs vary---ranging from models with up to 3 billion, 7 billion, or even 24 billion parameters---we adopt a practical perspective: we define SLMs as models capable of running efficiently on portable, resource-constrained devices such as smartphones. Under current hardware and optimization trends, this typically includes models around 3 billion parameters and, in some cases, up to 7 billion.

\begin{figure}
    \centering
    \includegraphics[width=1\linewidth]{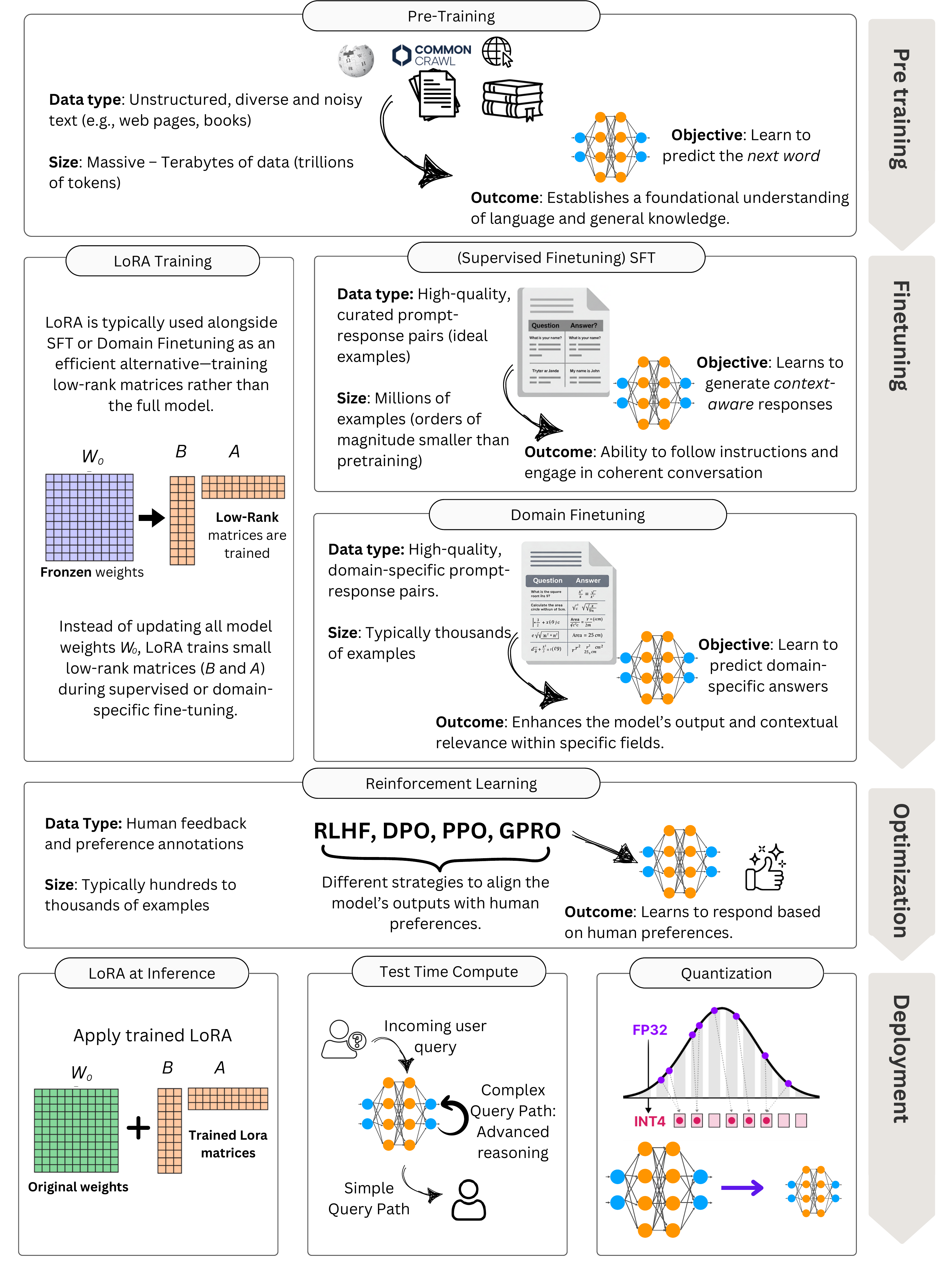}
    \caption{Overview of the training and deployment pipeline for large language models, illustrating the stages from pretraining to real-world application. Some steps are optional.}
    \label{fig:pipeline}
\end{figure}

Unlike their larger counterparts, which rely on extensive cloud-based infrastructure, SLMs emphasize low latency, cost-effectiveness, and ease of deployment, making them particularly well-suited for real-time, on-device AI applications. Recent benchmarking studies indicate that well-optimized SLMs can rival larger proprietary models in specific domains, all while requiring only a fraction of the memory and computational resources \cite{irugalbandara2024scalingscaleupcostbenefit, wang2024comprehensivesurveysmalllanguage}. This progress is fueled by advancements across multiple computational stages, including improved data curation, refined training methodologies, optimized architectural designs, enhanced fine-tuning strategies, and inference-time efficiency optimizations. As a result, SLMs are increasingly positioning themselves as practical, scalable solutions for AI applications that demand high performance without the burden of extensive infrastructure requirements. In the following sections, we delve into these advancements, exploring the key innovations that drive SLM efficiency, adaptability, and real-world applicability. Figure \ref{fig:pipeline} shows a major part of a pipeline for training.

\subsection*{Training-Time Techniques}
Pretraining forms the foundational stage where language models learn general linguistic patterns and world knowledge. While language models typically rely on massive web-scale datasets---often containing trillions of tokens from heterogeneous sources like websites, books, and online forums---this approach introduces significant noise and redundancy. SLMs adopt a more targeted paradigm, strategically optimizing data quality during pretraining to maximize capability within strict computational budgets. A key driver of SLM success is the use of high-quality, domain-specific training data, which allows these models to achieve competitive performance despite their smaller scale. By optimizing learning through targeted and relevant datasets, SLMs can effectively narrow the performance gap with larger models. Research shows that strategic data filtering—such as deduplication, domain selection, and dataset balancing—can significantly reduce the volume of required data while preserving generalization capabilities \cite{sajith2024trainingdataqualityquantity, lu2025smalllanguagemodelssurvey, wang2024comprehensivesurveysmalllanguage, kaddour2023minipilechallengedataefficientlanguage, shao2024deepseekmathpushinglimitsmathematical}. This strategy has proven particularly effective in domains like mathematical reasoning, where data quality can outweigh model size in determining performance \cite{guan2025rstarmathsmallllmsmaster}.

Architectural efficiency also plays a central role. As discussed earlier, the core component of transformer architectures is multi-head attention, which allows the model to analyze text from multiple perspectives simultaneously. However, this mechanism also introduces significant memory overhead. To mitigate these challenges, researchers have developed efficient attention mechanisms that optimize both computational and memory efficiency \cite{shazeer2019fasttransformerdecodingwritehead, ainslie2023gqatraininggeneralizedmultiquery, deepseekai2024deepseekv2strongeconomicalefficient, dao2022flashattentionfastmemoryefficientexact, gu2024mambalineartimesequencemodeling}. These methods achieve efficiency by either compressing stored representations, which capture word relationships within a sentence, or by enabling more effective sharing of information across attention heads. Such architectural improvements allow SLMs to process longer sequences while requiring significantly fewer hardware resources. Collectively, these innovations empower SLMs to deliver competitive performance at a fraction of the computational cost.

\subsection*{Fine-Tuning Techniques}
The transition from pretraining to real-world application is mediated through targeted fine-tuning strategies. 
While pretraining equips language models with broad linguistic and domain knowledge, fine-tuning tailors them to specialized applications. Instruction tuning, which involves training language models on conversational datasets using human-provided instructions, is essential for transforming pretrained models into responsive and interactive systems. Unlike foundation models focused on next-word prediction, instruction-tuned models learn to interpret context, follow multi-step commands, and generate pedagogically structured outputs. This makes models particularly well-suited for interactive applications such as AI-driven tutoring systems and specialized research assistants \cite{wang2024comprehensivesurveysmalllanguage}. 

The power of this approach is exemplified by DeepSeekMath, which employs instruction tuning alongside chain-of-thought and program-of-thought reasoning techniques. This specialized training strategy significantly improves the model's mathematical problem-solving abilities, illustrating how fine-tuning can optimize language models for complex cognitive tasks and domain-specific expertise \cite{shao2024deepseekmathpushinglimitsmathematical}.

Adapting SLMs to specialized domains such as science, mathematics, medicine, and legal reasoning is achieved through \textit{Domain Adaptation} \cite{bolton2024biomedlm27bparameterlanguage, yao2021adaptanddistilldevelopingsmallfast, yang2023mindllmpretraininglightweightlarge, Yang_2024, zhang2024sciinstructselfreflectiveinstructionannotated, zhang2024chemllmchemicallargelanguage, nguyen2023astrollamaspecializedfoundationmodels}. This process involves fine-tuning models on curated domain-specific datasets, enhancing their ability to understand specialized terminology and reasoning patterns. For example, mathematical reasoning models like DeepSeekMath undergo extensive pretraining on math-related corpora, significantly improving their problem-solving capabilities \cite{shao2024deepseekmathpushinglimitsmathematical}. By leveraging domain adaptation, SLMs can achieve performance comparable to or even surpassing larger models in specialized tasks, demonstrating the efficiency and practicality of fine-tuning for targeted applications.

\begin{table}[htbp]
    \small
    \centering
    \scalebox{0.9}{
    \begin{tabular}{l|ccc|ccc}
    \toprule
    \multirow{2}{*}{\textbf{Model}} & \multicolumn{3}{c|}{\textbf{English}} & \multicolumn{3}{c}{\textbf{Chinese}} \\
    \cmidrule{2-7}
    & GSM8K & MATH & MMLU-STEM & CMATH & Gaokao Cloze & Gaokao QA \\
    & \textit{8-shot} & \textit{4-shot} & \textit{4-shot} & \textit{6-shot} & \textit{5-shot} & \textit{4-shot} \\
    \midrule
    \multicolumn{7}{c}{\textit{General-Purpose Language Models}} \\
    \midrule
    LLaMA3.1-8B & 56.7 & 20.3 & 53.1 & 51.5 & 8.5 & 28.5 \\
    LLaMA3.1-70B & 85.5 & 41.4 & 78.1 & 75.5 & 11.9 & 43.3 \\
    LLaMA3.1-405B & 89.0 & 53.8 & -- & -- & -- & -- \\
    Qwen2-1.5B & 58.5 & 21.7 & 44.8 & 55.6 & 12.7 & 35.6 \\
    Qwen2-7B & 79.9 & 44.2 & 67.6 & 76.7 & 37.3 & 51.6 \\
    Qwen2-72B & 89.5 & 51.1 & 79.9 & 85.4 & 55.9 & 72.6 \\
    \midrule
    \multicolumn{7}{c}{\textit{Math-Tuned Specialized Models}} \\
    \midrule
    Qwen2-Math-1.5B & 71.3 & 44.4 & 50.4 & 79.6 & 37.3 & 50.7 \\
    Qwen2-Math-7B & 80.4 & 50.4 & 65.7 & 83.2 & 48.3 & 57.3 \\
    Qwen2-Math-72B & 89.1 & 60.5 & 79.1 & 86.4 & 72.9 & 69.5 \\
    \textbf{Qwen2.5-Math-1.5B} & 76.8 & 49.8 & 51.3 & 83.0 & 47.5 & 54.1 \\
    \textbf{Qwen2.5-Math-7B} & \textbf{91.6} & 55.4 & 67.8 & 85.0 & 57.6 & 69.5 \\
    \textbf{Qwen2.5-Math-72B} & 90.8 & \textbf{66.8} & \textbf{82.8} & \textbf{89.7} & \textbf{72.9} & \textbf{86.3} \\
    \bottomrule
    \end{tabular}}
    \caption{Performance of Qwen2.5-Math models versus general-purpose LLMs on mathematical reasoning benchmarks (GSM8K and MATH) under few-shot chain-of-thought prompting. Specialized smaller models such as Qwen2.5-Math-7B outperform much larger models like LLaMA3.1-405B, while even Qwen2.5-Math-1.5B, with only 1.5 billion parameters, surpasses LLaMA3.1-70B on the MATH benchmark \cite{yang2024qwen25mathtechnicalreportmathematical}.}
    \label{tab:qwen-math-table}
\end{table}

As an example of specialized fine-tuning in the mathematical domain, the Qwen2.5-Math series demonstrates how small, domain-adapted models can outperform larger general-purpose LLMs. As shown in Table~\ref{tab:qwen-math-table}, Qwen2.5-Math-7B surpasses LLaMA3.1-405B on GSM8K (8-shot) and MATH (4-shot) benchmarks, while the 1.5B variant outperforms LLaMA3.1-70B on MATH (4-shot). These results highlight the effectiveness of domain adaptation, showing that well-optimized SLMs can rival or exceed much larger models in reasoning tasks. This makes them especially promising for resource-constrained educational settings where math proficiency is a key goal.

A promising fine-tuning technique that improves the adaptability of large-scale language models while significantly reducing computational overhead is \textit{Low-Rank Adaptation (LoRA)} \cite{hu2021loralowrankadaptationlarge}. 
LoRA operates on the key observation that fine-tuning a pretrained model is inherently task-specific, 
meaning that the modifications required for adaptation primarily capture information relevant to a particular task rather than modifying the entire parameter space. Inspired by this, it is hypothesized that the weight updates during fine-tuning reside in a lower-dimensional subspace and thus exhibit a low intrinsic rank. Hence, instead of updating a full-rank weight matrix \( W_0 \in \mathbb{R}^{d \times k} \), LoRA approximates the update \( \Delta W \) using a low-rank decomposition:

\[
W' = W_0 + \Delta W = W_0 + BA
\]

where \( B \in \mathbb{R}^{d \times r} \), \( A \in \mathbb{R}^{r \times k} \), and the rank \( r \ll \min(d, k) \). Since task-specific adaptation affects only a small subset of the model's latent knowledge, the updates necessary for fine-tuning can be efficiently captured using low-rank matrices.

A key advantage of LoRA is its modularity. Unlike traditional fine-tuning approaches that modify all model parameters, LoRA freezes the pretrained model weights (\( W_0 \)), preserving general knowledge while introducing LoRA-specific layers (\( BA \)) to handle task-specific adjustments. This design enables multiple LoRA modules to be trained independently for different tasks and seamlessly integrated into the same base model as needed, analogous to swapping specialized tools on a multi-purpose device.

A useful analogy for understanding LoRA is a high-end camera with interchangeable lenses. The camera body (the pretrained model) remains unchanged, but different lenses (LoRA modules) can be attached depending on whether the user needs to capture landscapes, portraits, or low-light scenes. This plugin-like flexibility allows a single model to adapt to diverse tasks without requiring full retraining. By integrating low-rank trainable matrices while keeping the original model weights intact, LoRA dramatically reduces memory consumption and computational costs, often by orders of magnitude without sacrificing performance \cite{hu2021loralowrankadaptationlarge}. This efficiency makes LoRA particularly valuable for domain adaptation, multi-task learning, and personalized AI applications, enabling the deployment of multiple task-specific models without requiring extensive hardware resources.

Also ``Reinforcement learning-based fine-tuning" has been instrumental in refining model outputs based on human preferences and task-specific feedback, leading to significant improvements in language model optimization \cite{ouyang2022traininglanguagemodelsfollow, rafailov2024directpreferenceoptimizationlanguage, schulman2017proximalpolicyoptimizationalgorithms,shao2024deepseekmathpushinglimitsmathematical}. In particular, ``Group Relative Policy Optimization" has gained particular attention for its effectiveness in mathematical reasoning tasks, demonstrating substantial improvements in problem-solving accuracy and model adaptability \cite{shao2024deepseekmathpushinglimitsmathematical}. 

\subsection*{Test-Time Compute}

Test-time computing (TTC) plays a crucial role in enhancing the reasoning and performance of language models by dynamically adjusting computational effort during inference. This concept aligns with the cognitive framework described in the book ``Thinking, Fast and Slow" \cite{kahneman2011thinking}, which distinguishes between two modes of thinking: System 1, which is fast, intuitive, and effortless, and System 2, which is slower, deliberate, and cognitively demanding. Similarly, modern language models equipped with adaptive inference mechanisms balance efficiency and accuracy by selectively employing varying levels of test-time computation. Models configured for low TTC prioritize speed, generating rapid responses akin to System 1 thinking and it is sufficient for straightforward tasks but lacking the depth required for complex reasoning. Models with high TTC allocate more computational resources for deeper reasoning and improved accuracy. For example, answering factual questions like \textit{What is the capital of France?} might only require low-TTC processing, whereas solving multi-step mathematical proofs demands high-TTC deliberation to ensure accuracy. 

This adaptive strategy allows SLMs to adjust their resource usage based on the complexity of each task. It prioritizes speed and efficiency for simpler queries while allocating more computational effort to handle more challenging problems.~\cite{snell2024testtimecompute,liu20251bllmsurpass405b, snell2024scalingllmtesttimecompute}. Remarkably, studies show that small models with only 3~billion parameters optimized for adaptive TTC can outperform models 100~times larger on tasks like complex mathematical reasoning. For example, a 7~billion-parameter SLM leveraging TTC strategies has matched or surpassed leading commercial models like GPT-4o in specialized domains. Crucially, these methods reduce total computational costs by 100 to 1000× compared to traditional scaling, bypassing the need for massive pretraining~\cite{liu20251bllmsurpass405b}. By making high-performance AI viable on low-resource devices, TTC is emerging as a cornerstone of sustainable, accessible language technologies, enabling smaller models to deliver ``big model" capabilities at a fraction of the energy and cost.

\subsection*{Knowledge Distillation}

\begin{figure}[htbp]
    \centering
    \includegraphics[width=0.8\textwidth]{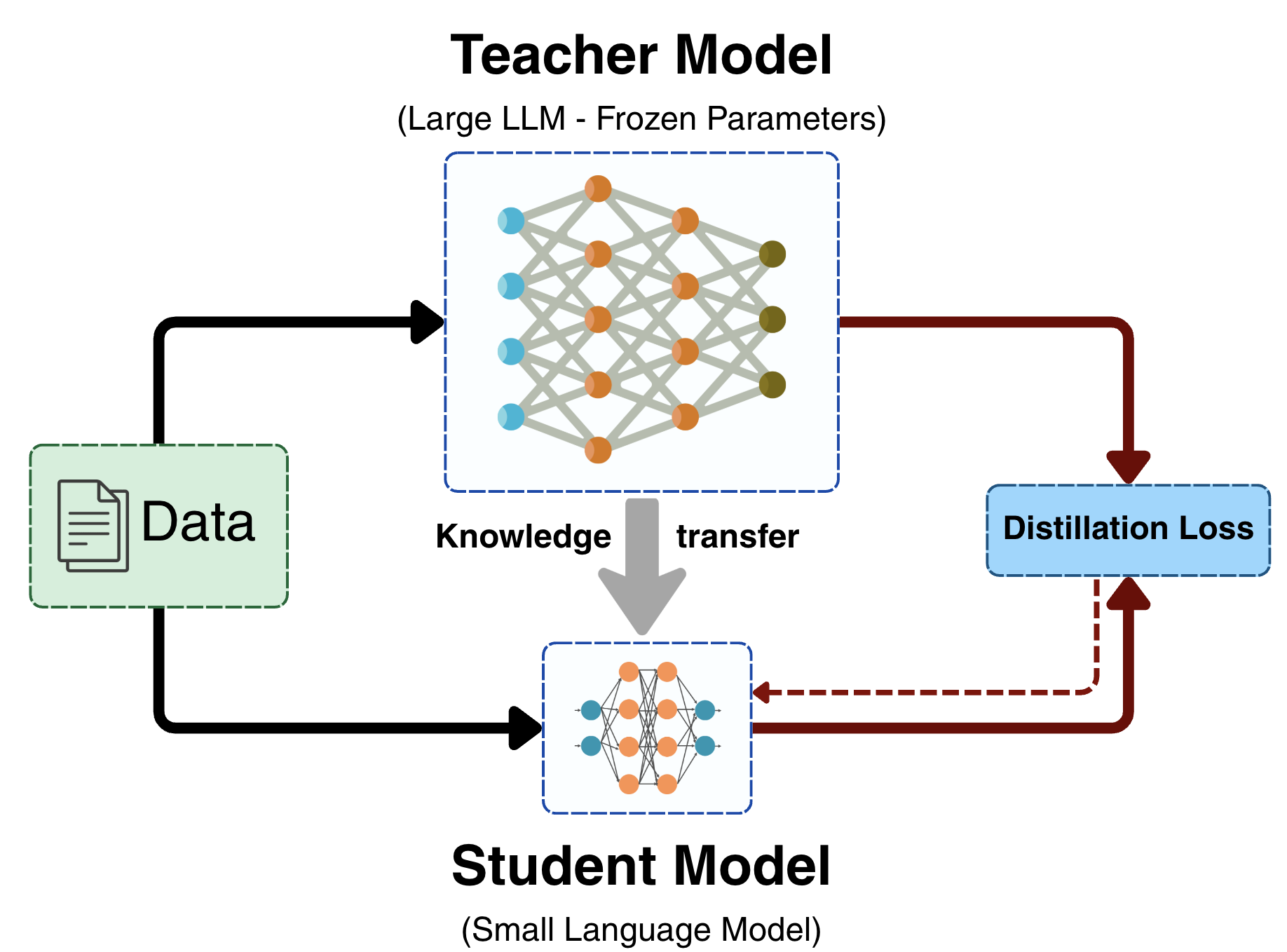}
    \caption{Illustration of the knowledge distillation process for training a student language model. A teacher LLM is guided by seed knowledge and skill-specific prompts to generate domain-relevant content. This generated knowledge is then used to train the student model according to a defined learning objective, enabling efficient model compression and targeted performance transfer \cite{xu2024surveyknowledgedistillationlarge}.}
    \label{fig:kd}
\end{figure}

Knowledge distillation enables the transfer of knowledge from large teacher models to smaller, more efficient student models, making it a crucial technique for enhancing performance, compressing models, and enabling self-improvement (see Fig. \ref{fig:kd}) \cite{xu2024surveyknowledgedistillationlarge}. Beyond simple output imitation, knowledge distillation facilitates the distillation of nuanced reasoning patterns and task-specific expertise \cite{chiang2023vicuna, wen2023fdivergenceminimizationsequencelevelknowledge, bai2022constitutionalaiharmlessnessai, yuan2024selfrewardinglanguagemodels}. This process allows smaller models to acquire advanced capabilities, such as instruction following and chain-of-thought reasoning, while maintaining computational efficiency \cite{mukherjee2023orcaprogressivelearningcomplex}. By transferring both explicit knowledge and latent reasoning structures, knowledge distillation significantly reduces computational costs, enabling smaller models to match or even surpass much larger counterparts. Recent studies show that compute-efficient distillation strategies can enable a 3 billion parameter (3B) model to outperform models over 100 times larger, demonstrating the power of well-optimized distillation techniques \cite{xu2024surveyknowledgedistillationlarge}.

Furthermore, self-distillation, a process in which models iteratively improve their outputs without the need for external teacher models, is gaining traction as a compelling alternative \cite{wang2023selfinstructaligninglanguagemodels}. As research progresses, optimizing self-improvement mechanisms, fine-grained alignment strategies, and vertical adaptation for specialized domains \cite{zhang2024sciinstructselfreflectiveinstructionannotated} will be essential for maximizing the efficiency and effectiveness of distilled models, further pushing the boundaries of scalable AI development.

Test-time compute and knowledge distillation empower small models to deliver high-quality reasoning previously associated only with much larger LLMs. As illustrated in Table~\ref{fig:deepseek-distill}, distilled variants of DeepSeek-R1-Distill outperform much larger non-reasoning models such as GPT-4o of OpenAI on a range of benchmarks. This highlights how efficient training and inference strategies can unlock advanced reasoning capabilities in compact models.

\begin{table}[h]
    \centering
    \resizebox{\linewidth}{!}{
    \begin{tabular}{@{}l *{6}{c} @{}}
    \toprule
    \multirow{3}{*}{\textbf{Model}} & \multicolumn{2}{c}{\multirow{2}{*}{\textbf{AIME 2024}}} & \multirow{2}{*}{\textbf{MATH-500}} & \textbf{GPQA} & \textbf{LiveCode} & \multirow{2}{*}{\textbf{CodeForces}} \\
    & & & & \textbf{Diamond} & \textbf{Benchmark} \\
    \cmidrule(lr){2-3}
    & pass@1 & cons@64 & pass@1 & pass@1 & pass@1 & rating \\
    \midrule
    GPT-4o (0513) & 9.3 & 13.4 & 74.6 & 49.9 & 32.9 & 759 \\
    Claude 3.5 Sonnet (1022) & 16.0 & 26.7 & 78.3 & 65.0 & 38.9 & 717 \\
    \midrule
    Distill-Qwen-1.5B (DeepSeek-R1) & 28.9 & 52.7 & 83.9 & 33.8 & 16.9 & 954 \\
    Distill-Qwen-7B (DeepSeek-R1) & 55.5 & 83.3 & 92.8 & 49.1 & 37.6 & 1189 \\
    Distill-Qwen-14B (DeepSeek-R1) & 69.7 & 80.0 & 93.9 & 59.1 & 53.1 & 1481 \\
    Distill-Qwen-32B (DeepSeek-R1) & \textbf{72.6} & 83.3 & 94.3 & 62.1 & 57.2 & 1691 \\
    Distill-LLaMA-8B (DeepSeek-R1) & 50.4 & 80.0 & 89.1 & 49.0 & 39.6 & 1205 \\
    Distill-LLaMA-70B (DeepSeek-R1) & 70.0 & \textbf{86.7} & \textbf{94.5} & \textbf{65.2} & \textbf{57.5} & 1633 \\
    \bottomrule
    \end{tabular}
    }
    \caption{Comparison of DeepSeek-R1 distilled models and other (non-reasoning) models. Despite their smaller size, the distilled variants of DeepSeek-R1 outperform or match significantly larger models like GPT-4o \cite{deepseekai2025deepseekr1incentivizingreasoningcapability}.}
    \label{fig:deepseek-distill}
\end{table}

\subsection*{Quantization}

Quantization is a method used to make LLMs faster and more efficient by reducing how much computer memory and processing power they need \cite{lang2024comprehensivestudyquantizationtechniques}. Normally, these models use high-precision numbers to represent the internal data they work with---like weights and activations. These values are typically stored in formats such as FP32 (32-bit floating point) or FP16 (16-bit), consuming 4 or 2 bytes of memory per parameter, respectively. With billions of such parameters, LLMs demand substantial memory and computational resources. Quantization addresses this challenge by converting model parameters into simpler, lower-precision formats such as INT8, INT4, or even INT2, using just 1 byte, half a byte, or a quarter of a byte per value. This compression can shrink model size by approximately 4×, 8×, or 16×, depending on the chosen format. While these lower-precision numbers are less accurate, quantized models often maintain strong performance across many tasks. As a result, quantization enables large models to run efficiently on smaller, less powerful devices like smartphones, laptops, or single-GPU systems rather than relying on expensive, high-end servers. Additionally, it accelerates computation by speeding up mathematical operations and reducing memory bandwidth requirements.

\section{Advancements in Inference and Deployment Frameworks for On-Device SLMs}

In addition to innovations in model architecture and training, a parallel evolution across hardware and software infrastructure has made it feasible to deploy SLMs directly on devices. Indeed, the true potential of SLMs emerges through their deployment in everyday devices like smartphones, laptops, and Internet of Things (IoT) systems, where they deliver AI capabilities without relying on cloud infrastructure. This transformation has been enabled by simultaneous innovations and advancements in three critical layers of the technology stack, working together to overcome the traditional barriers of power consumption, memory constraints, and computational complexity. In the following, we outline this stack in three conceptual tiers: hardware, frameworks, and applications.

\subsection*{Hardware Foundation: Specialized Chips for Everyday AI}
At the base of the AI software stack lies increasingly powerful mobile hardware, built around Systems-on-Chip (SoCs), a single integrated circuit that integrates multiple processor types onto a single chip. These SoCs are increasingly engineered to run language models and other machine learning workloads efficiently on-device, reflecting a broader shift toward AI-first hardware design.

Modern SoCs typically include three types of compute units. Central Processing Units (CPUs) handle general-purpose logic and control flow. Graphics Processing Units (GPUs) are optimized for parallel operations such as the matrix computations used in text generation. Neural Processing Units (NPUs) are purpose-built for high-throughput, low-power AI tasks. Prominent examples include Qualcomm's Snapdragon series, which feature Adreno GPUs and Hexagon NPUs; Apple's Neural Engine, which is integrated into Apple's A-series and M-series chips; and Google's Tensor SoC, custom-designed to accelerate AI tasks in Pixel devices. While there is still significant progress to be made, as language models continue to consume large amounts of power, these architectures are steadily improving the speed and energy efficiency of AI workloads. This enables more responsive, always-available AI assistants without putting a heavy strain on battery life~\cite{xu2024device}.

\subsection*{Framework Layer: Bridging Models to Hardware}

Inference frameworks are a crucial software layer that connects AI models to the underlying hardware. In this context, inference refers to the process of running a trained language model to generate outputs based on new input. These frameworks play a key role in optimizing AI workloads for edge deployment, ensuring that models run efficiently on a wide range of devices. These frameworks act as a translation layer, adapting AI models into forms that can take advantage of the specific capabilities of each device. To achieve this, inference frameworks handle several critical tasks. For example, they support the execution of quantized models—a technique discussed in the previous chapter as a way to reduce memory usage and computational demands. They also apply hardware-specific optimizations, tailoring operations to leverage the strengths of different processors such as CPUs, GPUs, or NPUs. In addition, they manage system resources by scheduling and distributing workloads to prevent bottlenecks on devices with limited power and capacity.

One notable example is \textit{llama.cpp}, a lightweight C and C++ library that provides an efficient foundation for LLM inference. The library supports a wide range of platforms, including mobile CPUs and GPUs, which makes it a popular choice for developers building AI applications on edge devices ~\cite{llamacpp2023}. Additionally, \textit{MLC-LLM} automatically converts language models into optimized code tailored to different hardware backends~\cite{mlc-llm}. This allows the models to run efficiently across a variety of devices. Another example is \textit{MNN}, an inference engine developed by Alibaba. It is specifically designed for mobile devices and includes features such as runtime optimization and backend abstraction. These capabilities enable it to support a wide range of hardware, including CPUs, GPUs, and NPUs, as well as different operating systems \cite{jiang2020mnnuniversalefficientinference}.

More recent projects such as \textit{PowerInfer-2}, \textit{HeteroLLM}, and \textit{llm.npu}, continue to push the boundaries of on-device AI by integrating advanced scheduling techniques and architectural strategies. These efforts aim to extract even more performance from mobile hardware~\cite{xue2024powerinfer2fastlargelanguage, chen2025heterollmacceleratinglargelanguage, xu2024fastondevicellminference}. Together, these toolchains play a crucial role in reducing the resource requirements for running AI models directly on devices, particularly those with limited memory and processing power.

\subsection*{Application Layer: Real-World On-Device SLM Use Cases}

At the top of the stack are real-world applications that integrate SLMs into user experiences across mobile and edge environments. Examples include \textit{Gemini Nano} on Google Pixel devices, which powers features such as smart replies, text summarization, and image captioning; \textit{Apple Intelligence}, which enables text rewriting and proactive task suggestions; and third-party apps like \textit{PocketPal AI}~\cite{pocketpalai2024}, which runs entirely offline using models such as Phi, Gemma, LLaMA, Qwen, and SmolLM. Tools like \textit{LM Studio} also make it possible for users to run models locally on desktops and laptops ~\cite{lmstudio}. These examples represent just a fraction of a rapidly expanding ecosystem of on-device language model applications that deliver private, low-latency AI capabilities without relying on the cloud. 

The on-device deployment of SLMs is made possible by a multi-layered system architecture. This stack begins with hardware acceleration, builds upon efficient software frameworks, and ultimately enables real-world applications. Each layer plays a vital role in making edge-based AI more accessible, private, and efficient.

\section{Applications of SLMs in Scientific and Technical Fields}

The recent advancements in SLMs have opened up numerous opportunities for scientific applications by enhancing efficiency, reducing computational costs, and improving adaptability across specialized domains. In mathematical reasoning, models like \textit{Llemma} have been fine-tuned using mathematical datasets such as \textit{Proof-Pile-2}, enabling SLMs to achieve strong performance on benchmarks like MATH and SAT, surpassing previous open-weight models \cite{wang2024comprehensivesurveysmalllanguage}. Similarly, models like \textit{DeepSeekMath} have demonstrated enhanced reasoning capabilities through reinforcement learning techniques \cite{shao2024deepseekmathpushinglimitsmathematical}. In scientific reasoning, SLMs such as \textit{SciGLM} employ self-reflective instruction annotation, allowing for collegiate-level reasoning across physics, chemistry, and other scientific disciplines \cite{zhang2024sciinstructselfreflectiveinstructionannotated, zhang2024chemllmchemicallargelanguage}. Additionally, \textit{AstroLLaMA}, fine-tuned on astronomy literature, has exhibited domain adaptability, improving tasks like automated paper summarization \cite{nguyen2023astrollamaspecializedfoundationmodels}.

In healthcare, specialized SLMs such as \textit{Hippocrates} have been designed for medical applications, achieving proficiency comparable to significantly larger models while benefiting from instruction tuning and reinforcement learning. Similarly, \textit{BioMedLM} has been trained exclusively on biomedical literature, demonstrating strong results in medical question-answering tasks \cite{bolton2024biomedlm27bparameterlanguage}. In coding, SLMs such as the \textit{Phi} series optimize computational efficiency and inference speed while maintaining competitive coding task performance. Beyond these domains, finance and legal applications are also benefiting from domain-specific SLMs like \textit{MindLLM}, which has been fine-tuned on legal and financial data to provide accurate and specialized decision support. The integration of domain-specific fine-tuning, reinforcement learning, and structured knowledge injection is enabling SLMs to rival much larger models in domain-specific tasks, positioning them as transformative tools in scientific and professional fields.

\begin{figure}
    \centering
    \includegraphics[width=1\linewidth]{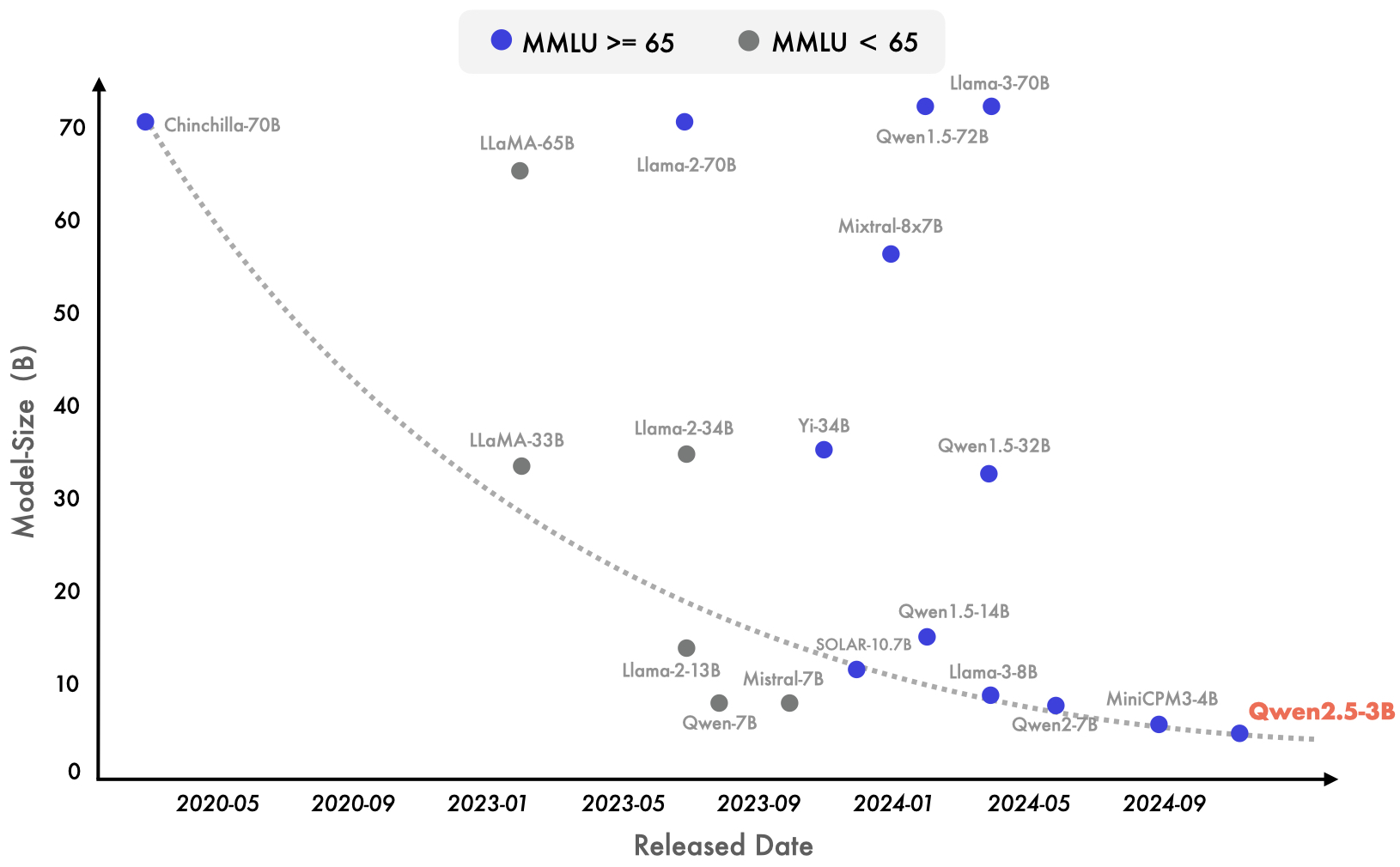}
    \caption{Performance of language models on the MMLU benchmark (higher is better). Recent SLMs, such as Qwen2.5-3B, show that models with around 3B parameters can exceed a score of 65, underscoring rapid gains in knowledge density and efficiency \cite{qwen2.5}.}
    \label{fig:mmlu-stem-performance-qwen2.5-small}
\end{figure}

Recent benchmarking results also demonstrate that modern SLMs are achieving performance levels on par with or even exceeding those of earlier-generation LLMs, particularly in STEM-related tasks (see Figure~\ref{fig:mmlu-stem-performance-qwen2.5-small}). Massive Multitask Language Understanding (MMLU) is a benchmark covering 57 academic and professional subjects, designed to assess reasoning and knowledge in a multiple-choice format \cite{hendrycks2021measuringmassivemultitasklanguage}. Recent evaluations on the MMLU benchmark, which includes subjects such as mathematics, physics, and chemistry, show that well optimized Small Language Models (SLMs) such as Qwen 2.5 and DeepSeekMath outperform many legacy large language models (LLMs) while remaining lightweight enough for deployment on edge devices \cite{qwen2025qwen25technicalreport, shao2024deepseekmathpushinglimitsmathematical}. This is a significant development for education in resource-constrained environments, as it indicates that learners can access high-quality reasoning capabilities without relying on expensive infrastructure.

\section{A Vision for Equitable and Inclusive Education}

Physics is at the heart of numerous technological advancements, including photonics, communication systems, renewable energy, and emerging quantum technologies \cite{nas2012optics}. However, effective teaching of these subjects requires more than traditional textbooks. It requires dynamic, interactive resources, opportunities for hands-on problem-solving, and tools that can demystify abstract concepts. In many developing countries, these needs are compounded by a lack of educational infrastructure and insufficient political support for science education \cite{Talisayon2002}. 

SLMs, unlike their large-scale counterparts, are optimized to operate efficiently on low-power devices with limited computational capacity. They can function offline on laptops, smartphones, or portable servers, removing the dependency on stable internet connectivity. This makes them especially valuable in educational settings where internet access is inconsistent or cost-prohibitive. SLMs hold the promise to bring several notable advantages to physics and photonics education. They have the capacity to enable interactive learning by serving as on-demand, offline virtual tutors. Students can receive immediate, context-specific explanations of complex topics, such as Maxwell’s equations or the principles of optical coherence, enhancing their understanding and engagement. SLMs can be fine-tuned for language localization, allowing them to operate in students’ native languages. This feature helps overcome linguistic barriers that frequently hinder STEM education. For instance, adapting scientific content to local languages across African regions has proven to increase accessibility and foster inclusive learning environments \cite{bamgbose2021mother, kamwangamalu2022african}. 

Educators also stand to benefit from SLMs. These models can assist in curriculum design by generating lesson plans, creating problem sets, and translating dense academic texts into simpler language. Teachers in remote or under-resourced schools, who often lack access to peer collaboration or up-to-date materials, gain a flexible assistant capable of supporting their instructional needs \cite{choi2024llms, huber2024leveraging}. Moreover, by offering culturally and ethically adaptive outputs, SLMs can align educational content with local values and social contexts—an essential consideration for meaningful and sustainable learning \cite{cipit2021ethical}. 

While SLMs offer exciting potential, it is important to acknowledge their current limitations---many of which remain active areas of research and must be considered when implementing real-world solutions. Hallucination remains a key concern, particularly in educational contexts where factual accuracy is critical. Additionally, multilingual performance still lags in many low-resource languages~\cite{allal2025smollm2smolgoesbig}, limiting accessibility for linguistically diverse communities.

Perhaps most powerfully, SLMs can foster collaborative and project-based learning. Students can use shared devices to work together on science problems, explore simulations, or conduct guided discussions, with the model acting as a moderator or knowledge source. In settings where lab access is limited or nonexistent, this kind of digitally mediated collaboration becomes a valuable substitute for hands-on experimentation \cite{rao2004physics}. Realizing this vision, however, will require collective action to support the development and deployment of localized AI technologies. Investments in affordable hardware are essential to ensure that these tools reach the classrooms and communities that need them most \cite{worldbank2024ai}.

Beyond education, SLMs could also address broader structural challenges associated with cloud-based AI. They reduce reliance on expensive, centralized infrastructure, mitigate latency issues and environmental costs, and enhance data privacy by keeping sensitive information on-device \cite{irugalbandara2024scalingscaleupcostbenefit, theverge2020awsoutage, epa_greenhouse_gas, touvron2023llama2openfoundation, grattafiori2024llama3herdmodels}. These qualities are particularly relevant in sectors like healthcare, defense, and field research, where network access is often unreliable, and confidentiality is paramount.

Small language models hold immense potential to bridge the educational gap and digital divide in underdeveloped regions, providing students and educators with tools to explore the wonders of physics and photonics without relying on internet access. By harnessing the capabilities of offline AI solutions, we can create a future where every student, regardless of their circumstances, has the opportunity to engage with cutting-edge knowledge and pursue their academic dreams. As educators and researchers, it is our collective responsibility to ensure that no one is left behind in the quest for scientific discovery and innovation.

\section*{Acknolawdgement}
The authors gratefully acknowledge Loubna Ben Allal for her valuable feedback on the manuscript.

\bibliography{main.bib}
\end{document}